\documentclass[12pt]{article}
\usepackage{amssymb}
\usepackage{a4wide}
\begin{document}

{\renewcommand{\thefootnote}{\fnsymbol{footnote}}
%\hfill  IGC--11/12--2\\
\medskip
\begin{center}
{\LARGE  A no-singularity scenario in loop quantum gravity}\\
\vspace{1.5em}
Martin Bojowald\footnote{e-mail address: {\tt bojowald@gravity.psu.edu}}
 and George M.\ Paily\footnote{e-mail address: {\tt gmpaily@phys.psu.edu}}
\\
\vspace{0.5em}
Institute for Gravitation and the Cosmos,
The Pennsylvania State
University,\\
104 Davey Lab, University Park, PA 16802, USA\\
\vspace{1.5em}
\end{center}
}

\setcounter{footnote}{0}

\newcommand{\lP}{\ell_{\mathrm P}}
\newcommand{\md}{{\mathrm{d}}}
\newcommand*{\R}{{\mathbb R}}
\newcommand*{\N}{{\mathbb N}}
\newcommand*{\Z}{{\mathbb Z}}
\newcommand*{\Q}{{\mathbb Q}}
\newcommand*{\C}{{\mathbb C}}

\begin{abstract}
  Canonical methods allow the derivation of effective gravitational actions
  from the behavior of space-time deformations reflecting general
  covariance. With quantum effects, the deformations and correspondingly the
  effective actions change, revealing dynamical implications of quantum
  corrections. A new systematic way of expanding these actions is introduced
  showing as a first result that inverse-triad corrections of loop quantum
  gravity simplify the asymptotic dynamics near a spacelike collapse
  singularity. By generic quantum effects, the singularity is removed.
\end{abstract}

\section{Introduction}

Quantum gravity, with general relativity as its limit at small
gravitational potential and curvature, endows space-time with quantum
effects. Depending on the approach used, there are different
modifications to the classical dynamical equations. It is difficult to
analyze these theories directly except in model systems which often
focus on just one type of modification and severely restrict the class
of solutions looked for, for instance by symmetry reduction. With such
restrictions, on the other hand, it is impossible to say how generic
the results are. Here, we employ and extend a new application of an
old technique in canonical gravity to evaluate {\em unrestricted} loop
quantum gravity regarding one important issue: the singularity
problem.

\section{Singularities}

The singularity problem of general relativity states that space-time is
generically incomplete, with only finite ranges of time over which we can
extrapolate in some directions. The most infamous example, the big-bang
singularity, shows that general relativity does not allow us to extend our
solutions to space-time before the big bang. As in this case, singularities
often, but not always, are signaled by diverging curvature quantities or, in
physical terms, infinite tidal forces and energy densities. It is possible to
evade singularities classically \cite{NonSingSol}, but no general mechanism is
known to avoid the strict verdict of singularity theorems without fine-tuning.

At large curvature, where many cases of singularities occur, quantum gravity
is expected to be important. It is therefore a common hope that quantum
effects, for instance new repulsive contributions to the gravitational force,
can help to avoid singularities generically; indications do indeed exist
\cite{Sing,HomCosmo} in reduced models especially of loop quantum cosmology
\cite{LivRev,Springer}. The general mechanism shows that wave functions for
the universe, subject to difference equations \cite{cosmoIV}, can be extended
through a classical singularity. A more intuitive picture, available in some
cases, is a ``bounce'' \cite{GenericBounce,QuantumBigBang,BounceReview} of the
cosmological scale factor when the energy density is nearly
Planckian.\footnote{Generically, bounces in loop quantum cosmology are
  realized with matter whose energy is dominated by its kinetic contribution
  \cite{BouncePert}. For other matter ingredients it is not clear how
  generically bounces occur since quantum back-reaction --- or
  higher-curvature corrections --- remains poorly controlled in those cases
  \cite{BounceSqueezed}.}  This effect, however, makes use of only one quantum
effect, called holonomy correction as introduced in more detail below, and
implicitly or explicitly ignores several others. Holonomy corrections in
isotropic cosmological models bound the energy density $\rho$ from above by a
nearly Planckian upper limit $\rho_{\rm max}$, and correspondingly are
strongest when $\rho/\rho_{\rm max}$, or $\ell_{\rm P}^2/\ell_{\cal H}^2$ with
the Hubble distance $\ell_{\cal H}=1/{\cal H}$, is of the order one. The same
size of corrections is expected from higher-curvature terms (or loop diagrams
in perturbative quantum gravity \cite{EffectiveGR}), but these are ignored in
most bounce models of loop quantum cosmology. These models, therefore, do not
provide generic scenarios.

Moreover, it is non-trivial to extend bounce scenarios to
inhomogeneous geometries.  Regarding generic inhomogeneous
singularities, one often tries to appeal to the
Belinskii--Khalatnikov--Lifschitz (BKL) conjecture \cite{BKL}: Near a
spacelike curvature singularity, time derivatives seem to become
dominant in the dynamics of general relativity. (Spatial derivatives
can be large, but they do not significantly contribute to the
evolution.) An expanding or collapsing universe is then essentially a
collection of homogeneous cosmological models, one per spatial
point. Homogeneous models are much easier to analyze, especially in
loop quantum cosmology \cite{HomCosmo}. If they can be shown to become
non-singular by quantum effects, the full BKL-like dynamics may be
non-singular.

The applicability of these ideas depends on how singularities are
resolved.  The asymptotic nature of BKL arguments and the definitive
turning point of a bounce do not agree well, making singularity
resolution in inhomogeneous situations difficult. Moreover, despite
some progress \cite{PastAttract,GeneralHarmonic,WS:AR} the BKL
scenario remains to be proven and fully formulated even
classically. Some difficulties concern the fact that the scenario is
not obvious from the action or equations of motion but rather follows
from an often painstaking analysis of solutions. This fact makes
results depend on a preferred time variable, derivatives by which
become dominant. Such a choice of time might be natural for an
analysis of the present, nearly homogeneous universe, but becomes
questionable when assumed near a singularity. It remains unclear whether
the BKL scenario can be compatible with general covariance.
%In this article, we demonstrate that specific features of loop quantum
%gravity, derived in general terms using effective techniques, provide
%a natural substitute of the BKL scenario regarding its generality and
%covariance.

\section{Off-shell loop quantum gravity}

The dependence on what is considered time, or a spatial slice at constant
time, points to another important problem, the recognition of which will
eventually allow us to make progress. In a canonical formulation of general
relativity, space-time covariance is implemented by local symmetries under
deformations of spatial slices in space-time \cite{DiracHamGR}. There are
three independent deformations generated by $D[N^i]=\int{\rm d}^3x
N^i(x)D_i(x)$ tangential to space and one, $H[M]=\int{\rm d}^3x M(x)H(x)$,
normal to space, with $D_i(x)$ and $H(x)$ depending on the spatial metric
$g_{ij}$ and its rate of change. Geometrical considerations of deformations
imply that the symmetry generators must satisfy a certain algebra under
Poisson brackets; in particular, for $H[M]$ we have the relationship
\begin{equation} \label{HH}
 \{H[M_1],H[M_2]\}= D[\beta(M_1\vec{\nabla} M_2-M_2\vec{\nabla}M_1)]
\end{equation}
with $\beta=1$ classically. That an algebra of this form be realized
is not only a crucial condition for the consistency of any space-time
theory, including quantum gravity perhaps with quantum corrections
$\beta\not=1$; Eq.~(\ref{HH}) is also a key tool for an analysis of
the dynamics, for it contains much information: The classical
equations of motion (to second derivative order) follow from it
\cite{Regained,LagrangianRegained}, and higher derivative orders are
restricted to ensure covariance. 

However, an analysis of relations such as (\ref{HH}) --- and as a
consequence the general problem of defining a consistent quantum
theory of gravity --- is usually complicated by the fact that not only
structure constants but also phase-space functions enter, in
particular the spatial metric used to define the gradient
$\nabla^i=g^{ij} \partial/\partial x^j$.  In particular, it is not
known whether loop quantum gravity \cite{Rov,ALRev,ThomasRev} is fully
consistent, but it has led to consistent deformations (\ref{HH}) with
$\beta\not=1$ in several model systems and for quantum corrections of
different types. 

Loop quantum gravity has a dynamics of SU(2)-Yang--Mills form, with a
connection $A_j^I$ and a conjugate field, the densitized triad
$E^j_I$.\footnote{To be closer to the notation of
\cite{Regained,LagrangianRegained} used crucially for some
derivations, we deviate from the usual practice in loop quantum
gravity of denoting tangent-space indices by $a,b,c,\ldots$. Those
indices are labelled $i,j,k,\ldots$, while $I,J,K,\ldots$ are used for
internal indices. We will suppress the Barbero--Immirzi parameter in
this article.} To quantize them, these fields are integrated along
curves and surfaces, respectively, to obtain holonomies of $A_j^I$ and
fluxes of $E^j_I$. Using a U(1)-simplification for the sake of
illustration, we write $A^I_j=c_{(j)}(x)\delta_j^I$ and
$E^j_I=p^{(j)}(x)\delta_j^I$. Integrated variables, labeled by curves
$e$ along which integrations are done, are
$c_e=\int_{e}t^jc_j(\lambda) {\rm d}\lambda$ and
$F_S=\int_{S}n_jp^j(y) {\rm d}^2y$ with curves $e$ tangent to $t^j$
and surfaces $S$ co-normal to $n_j$. Loop quantum gravity is based on
a representation where all holonomies $h_e=\exp(i c_e)$ act as shift
operators on a U(1)-theory per edge $e$, and fluxes $F_S$ act by
derivatives with discrete spectra \cite{AreaVol}.

The specific kinematics implies characteristic corrections on top of the
usually expected higher-curvature terms and higher time derivatives in quantum
dynamics (which in a canonical quantization follow from quantum back-reaction
\cite{EffAc,Karpacz}). The use of holonomies $h_e$ for $c_j(x)$ in Hamiltonian
operators and in $H[M]$ implies higher-order corrections by powers of $c_j(x)$
or $\dot{p}^j(x)$, and also higher spatial derivatives if a derivative
expansion for the integrated $c_e$ is used. Both consequences resemble terms
from higher-curvature corrections, but they are not the same because they lack
higher time derivatives. Accordingly, holonomy corrections, unlike
higher-curvature ones, modify the space-time structure by implying quantum
corrections in (\ref{HH}). Several examples for consistent deformations exist
in which $\beta$ is of the form $\cos(2\delta c_j(x))$ with a parameter
$\delta$ depending on the discreteness scale (the length of curves $e$ used to
construct a state). They include spherical symmetry \cite{LTBII,JR}, $2+1$
gravity \cite{ThreeDeform}, and perturbative inhomogeneity
\cite{ScalarHol}.

A second type of quantum-geometry corrections is more indirect: Matter
Hamiltonians and $H[M]$ always require inverses of $p^j(x)$, but the
quantized $F_S$ with their discrete spectra containing zero do not
have inverse operators. Instead, general techniques \cite{QSDI} exist
to derive operators with the inverse as their classical limit: Pick
intersecting pairs of curves $e$ and surfaces $S$, use Poisson
brackets $\{c_e,F_S\}=8\pi G$ with the gravitational constant $G$ and
write $|F_S|^{-1/2}{\rm sgn}(F_S)= \{c_e,|F_S|^{1/2}\}/4\pi G=
-i(h_e^*\{h_e,|F_S|^{1/2}\}- h_e\{h_e^*,|F_S|^{1/2}\})/8\pi G$. Loop
quantum gravity has quantized $\hat{h}_e$ as finite shift operators,
for which commutator identities show that $\widehat{|F_S|^{-1/2}{\rm
sgn}(F_S)}=(|\hat{F}_S+8\pi\ell_{\rm P}^2|^{1/2}-
|\hat{F}_S-8\pi\ell_{\rm P}^2|^{1/2})/8\pi\ell_{\rm P}^2$ is
well-defined (and zero) even on zero-eigenstates of $\hat{F}_S$ where
the classical inverse would diverge. At small flux values, these
constructions imply inverse-triad corrections \cite{InvScale} with
classical inverses such as $1/p^j$ replaced by $\alpha(p^i)/p^j$ for a
correction function $\alpha$ that vanishes at zero $p^j$. Consistent
deformations (\ref{HH}) with $\beta=\alpha^2$ have been found in
spherical symmetry \cite{LTBII,ModCollapse}, in $2+1$ gravity
\cite{TwoPlusOneDef}, and with perturbative inhomogeneity
\cite{ConstraintAlgebra}.\footnote{Also some Wheeler--DeWitt models
have been argued to give rise to deformed hypersurface-deformation
algebras \cite{BohmEuclidean}.} In some models, holonomy and
inverse-triad corrections have been combined, showing multiplicative
behavior of the deformation function $\beta$. With spherical symmetry,
for instance, the combined correction function is
$\beta=\alpha(p^i)^2\cos(2\delta c_j)$ \cite{JR}.  Several rather
different systems have been analyzed, with various methods (effective
techniques and operator calculations). In all cases, deformed algebras
of related forms have been found, while undeformed consistent versions
do not seem to exist.

The presence of all the corrections makes it hard to keep an overview
of the evolution. The situation is much clearer at the level of the
algebra (\ref{HH}), whose corrections contained in $\beta-1$ appear to
be rather universal. With canonical techniques, one can reconstruct an
effective action from the deformed algebra for $\beta\not=1$. An
expansion by the rate of change of the spatial metric then shows
general properties of loop quantum gravity that reveal the space-time
structure near a collapse singularity.

\section{Effective action}

To that end, as in \cite{LagrangianRegained}, we first perform a
Legendre transformation from the Hamiltonian $H(x)$, depending on $x$
via the spatial metric $g_{ij}$ and its momentum $\pi^{ij}$, to
$L(x)=\pi^{ij}v_{ij}-H(x)$ with $v_{ij}(x)=N^{-1}\delta H[N]/\delta
\pi^{ij}(x)$ the normal rate of change of $g_{ij}$. For derivatives by
the spatial metric, we have $\delta H/\delta g_{ij}= -\delta L/\delta
g_{ij}$. The relation (\ref{HH}), written explicitly with a Poisson
bracket $\{f,g\}= \int{\rm d}^3x (\delta f/\delta g_{ij}(x))(\delta
g/\delta \pi^{ij}(x))- (f\leftrightarrow g)$, then implies $
v_{ij}(x')\delta L(x)/\delta g_{ij}(x) + \beta
D^i\nabla_i\delta(x,x')- (x\leftrightarrow x')=0$.  A metric changes
under spatial deformations such that $D^i= -2\nabla_j\pi^{ij}$, and
after replacing $\pi^{ij}$ with $\delta L/\delta v_{ij}$ we obtain a
linear equation
\begin{equation} \label{L}
\frac{\delta L(x)}{\delta g_{ij}(x')} v_{ij}(x') \delta(x,x')-
 (x\leftrightarrow x') 
= -2\frac{\delta L}{\delta v_{ij}}\nabla_j\left(\beta
  \nabla_i\delta(x,x')\right)- (x\leftrightarrow x')\,. 
\end{equation}

We can solve this equation order by order in an expansion by $v_{ij}$, writing
\begin{eqnarray} \label{LK}
  L(x)&=& \sum_{n=0}^{\infty}
  \sum_{N_1,\ldots,N_n=0}^{\infty}
L^{(i_1,j_1,k_1^{(1)},\ldots,k_1^{(N_1)}),\ldots,
  (i_n,j_n,k_n^{(1)},\ldots,k_n^{(N_n)})}(g_{ij})\\
&&\times(\nabla_{k_1^{(1)}}\cdots\nabla_{k_1^{(N_1)}} v_{i_1j_1})\cdots 
(\nabla_{k_n^{(1)}}\cdots\nabla_{k_n^{(N_n)}} v_{i_nj_n})\nonumber
\end{eqnarray}
and an analogous expansion for $\beta(x)$.  To be specific, we will
assume that only even terms appear in (\ref{LK}), to ensure
time-reversal invariance. (If the assumption is violated, our
arguments still go through.)  We deviate from
\cite{LagrangianRegained} because of the presence of $\beta$ as well
as higher spatial derivatives in the expansion. The former is a
consequence of quantum space-time with a deformed relation (\ref{HH}),
and the latter allows for non-local effects in holonomies and
fluxes. Moreover, unlike \cite{LagrangianRegained} which derived the
classical action to second order in $v_{ij}$ and in derivatives, we
are interested in violent near-singular regimes in which many terms of
the expansion may be relevant.

We focus on $\beta^{\emptyset}(g_{ij})$, the leading coefficient in
the expansion of $\beta$ independent of $v_{ij}$. From the form of
consistent deformations known, this coefficient is sensitive to
inverse-triad corrections, and it becomes very small near a classical
spacelike collapse singularity where fluxes are zero
\cite{InvScale}. Coefficients of $\beta$ of higher order in the
$v$-expansion, on the other hand, are determined by all types of
quantum corrections and do not have characteristic values.

Solving (\ref{L}) order by order as a recurrence relation for coefficients in
(\ref{LK}) shows that terms of order $n+2$ in $v_{ij}$ and its
spatial derivatives are related to terms of lower order $n$ {\em divided by}
$\beta^{\emptyset}$: Take the coefficients of order $n+1$ in the $v$-expansion
of all terms in (\ref{L}), obtained by expanding $L$ and $\beta$ and
collecting terms of the same order in $v_{ij}$ and its spatial
derivatives. The highest-order $L$-coefficient, of order $n+2$, is obtained
from $v_{ij}$-derivative terms on the right-hand side with only the leading
coefficient $\beta^{\emptyset}$ of $\beta$ (or $\nabla_j\beta^{\emptyset}$
which too is small for small fluxes) as a factor. Higher orders in
$\beta$ are multiplied with $L$-coefficients of order at most $n$ so as to
produce a total order of $n+1$. On the left-hand side of the equation, order
$n+1$ is obtained with $L$-coefficients of order $n$. The $L$-coefficient of
order $n+2$ multiplied with $\beta^{\emptyset}$ then equals a combination of
$L$-coefficients of order at most $n$ and without a factor of
$\beta^{\emptyset}$.

To second order in the derivative expansion, for instance, we have the
effective action $L_2=(16\pi G)^{-1}\sqrt{\det g} \left({\rm
    sgn}(\beta^{\emptyset})|\beta^{\emptyset}|^{-1/2} {\cal
    G}^{ijkl}v_{ij}v_{kl}+ \sqrt{|\beta^{\emptyset}|}
  {}^{(3)}\!R-2\Lambda\right)$ with ${\cal G}^{ijkl}=\frac{1}{4}(g^{i(k}
g^{l)j}- g^{ij}g^{kl})$ and the spatial Ricci scalar ${}^{(3)}\!R$
\cite{Action}. Continuing the expansion to higher orders and using our
recurrence arguments then shows that a term of order $n$ in the $v$-expansion
is multiplied with $|\beta^{\emptyset}|^{(1-n)/2}$. Terms with higher powers
of $1/\beta^{\emptyset}$ play the largest role near a spacelike singularity,
while others can be ignored. But also the typical size of products of $v_{ij}$
and its spatial derivatives matters, which could compensate for suppression
factors of $\beta^{\emptyset}$.

\section{Derivative expansion}

In order to see what the dominant terms are generically, we combine the
expansion in $v_{ij}$ with a derivative expansion. The $v$-expansion by $n$ in
(\ref{LK}), in which we consider $v_{ij}$ and its spatial derivatives as being
of the same order, is suggested by the form of (\ref{L}) with its different
dependences on $v_{ij}$. The $v$-order is different from the derivative order,
which is useful to tell what terms are generically of similar orders in a
strong-curvature regime.  In contrast to the $v$-expansion, the derivative
expansion does not distinguish between space and time derivatives.  One factor
of $v_{ij}$ in our expansion contributes one derivative order because
$v_{ij}$, as the rate of change of $g_{ij}$, has one time
derivative. Including spatial derivatives, terms of fixed derivative order $N$
are of the form
\begin{eqnarray}
&& |\beta^{\emptyset}|^{(1-N)/2} v^N+
|\beta^{\emptyset}|^{(2-N)/2} (v^{N-1}g'+v^{N-2}v') \label{beta}\\
&&+
|\beta^{\emptyset}|^{(3-N)/2} \left(v^{N-2}(g''+(g')^2)
+ v^{N-3}(v''+v'g')+ v^{N-4}(v')^2\right)+\cdots \nonumber
\end{eqnarray}
where we do not spell out individual coefficients and indices. All terms have
the same number of derivatives, but different numbers of $v$-factors as
indicated by the powers of $\beta^{\emptyset}$. Terms of equal numbers of
derivatives are generically of the same size in a given curvature regime, and
factors of $\beta^{\emptyset}$ decide which terms are important for the
dynamics. For small $\beta^{\emptyset}$, the leading term,
$|\beta^{\emptyset}|^{(1-N)/2} v^N$, is dominant in the expansion (\ref{beta}),
and it does not contain any spatial derivatives.

Near a spacelike singularity, the effective action therefore contains
only time derivatives, obeying the dynamics of some homogeneous
model. The picture is reminiscent of the BKL scenario, but in the form
found here, making use of quantum effects, it has several
advantages. First, the dominance of time derivatives follows from a
consideration of the effective action, without the need to solve
equations of motion. Secondly, we do not make use of a preferred time
coordinate or spatial slicing of space-time. Our considerations are
covariant because they are based on a consistent deformation
(\ref{HH}) of the classical covariance algebra. It is the
quantum-corrected covariance algebra itself which implies the
suppression of spatial derivatives, not the choice of a preferred
slicing that would break covariance. Finally, our arguments remain
valid at a general level of perturbative quantum gravity, referring
to all orders in the curvature expansion.

\section{No singularities}

We are now ready to apply our new scenario to the singularity
problem. Imagine that we follow the evolution of some quantum geometry
toward a spacelike collapse singularity.  Effective equations describe
the dynamics of states parameterized by expectation values and
moments, as well as quantum-geometry corrections.  When curvature
grows large, different effects become important. They may stop the
collapse or trigger a bounce, in which case the singularity is not
reached.\footnote{As a further consequence of deformed algebras,
holonomy corrections do not imply a true bounce, as it seems in some
isotropic models \cite{QuantumBigBang}, but rather trigger signature
change since $\beta$ turns negative at near-Planckian curvature
\cite{Action}. If evolution does not stop by holonomy effects and
inverse-triad corrections become large, we always have
$\beta^{\emptyset}>0$.}  But if there is nothing to stop the approach
to vanishing fluxes, which classically amounts to a singularity,
inverse-triad corrections will make $\beta^{\emptyset}$ smaller and
smaller. By our new scenario, time derivatives in the effective action
then dominate and we can continue the evolution by using homogeneous
mini\-superspace models. At some point, effective actions will break
down or all terms in the derivative expansion will have to be
considered. But instead of doing this, we can use exact quantizations
of homogeneous models in loop quantum cosmology and conclude,
following \cite{Sing,HomCosmo}, that they are non-singular: any
quantum state is extended uniquely across the classical
singularity. Once we extend the state through vanishing fluxes,
effective actions will again be available and continue our evolution.

Generically, no collapse singularity occurs in loop quantum gravity, even for
geometries not required to obey any symmetry. The key mechanism is
inverse-triad corrections and the new space-time structure they imply via
(\ref{HH}).  These corrections both provide the suppression of spatial
derivatives in effective actions and facilitate the extension of wave
functions across a classical singularity.

Our treatment does not apply to all space-time metrics because one could
choose a time coordinate and initial values such that time and spatial
derivatives are unbalanced, with spatial ones unnaturally large compared to
time ones. However, this would require fine-tuning and provide a negligible
subset of universe models. And changing the space-time slicing would bring one
back to a situation in which our arguments do apply. Only the expansion is
more difficult to organize in some slicings; our mechanism of singularity
resolution, which is covariant, is always realized.

\section{Conclusions}

In spite of its generality, our scenario is not a complete proof of
the absence of singularities in loop quantum gravity, and we refrain
from calling it a ``no-singularity theorem.'' We must assume that loop
quantum gravity can be consistent, obeying local symmetries of a
covariance algebra of the form (\ref{HH}) with some function
$\beta$. Good evidence for this behavior exists from several model
systems
\cite{ConstraintAlgebra,JR,LTBII,ThreeDeform,ModCollapse,ScalarHol,TwoPlusOneDef},
and no undeformed off-shell algebra has been found, but a
demonstration in general terms seems currently out of reach. Our
arguments then show that any consistent realization is generically
non-singular. They also highlight the importance of off-shell
properties of canonical quantum gravity, whose daunting derivations
are often avoided by resorting to gauge-fixing or deparameterization
before quantization, leaving the quantum system contaminated with
potentially inconsistent gauge artefacts.

We have provided not only the first general statement about
non-singular evolution in quantum gravity; we have also shown how
dynamical consequences can be derived taking into account all expected
corrections of quantum gravity. Holonomy corrections and
higher-curvature corrections both contribute to higher orders in the
$v$-expansion and can have significant effects, but they leave the
leading coefficient $\beta^{\emptyset}$ unchanged and therefore do not
interfere with the resolution of singularities. For inverse-triad
corrections, we have explicitly displayed expressions obtained with a
U(1)-simplification. With the non-abelian SU(2), some expressions are
more complicated, providing an additional dependence on $v_{ij}$
\cite{DegFull}: holonomies in terms such as
$\hat{h}_e[\hat{h}_e^{\dagger},|\hat{F}_S|^{1/2}]$ no longer cancel
completely. Also these contributions can be taken care of by changing
higher orders in the $v$-expansion, without affecting
$\beta^{\emptyset}$.

In addition to solving the singularity problem in loop quantum gravity, our
methods have several other consequences. By relating inhomogeneous to
homogeneous evolution in general terms, they show how (and what)
minisuperspace effects can be used for general geometries, and how
reliable they are. We see that especially inverse-triad effects, with their
characteristic form, survive even in regimes in which curvature is large and
other quantum effects play equally important roles. We also provide a new
scenario for the approach of space-time toward a classical singularity, which
is similar in spirit to the BKL picture but relies crucially on quantum
effects and avoids several downsides of the BKL arguments. Given the influence
that the BKL scenario has had over the past few decades, we expect that our
new scenario will play an important role for further analysis of the dynamics
of loop quantum gravity.

%\ack

\bigskip

\noindent This work was supported by NSF grant PHY0748336.

%\bibliographystyle{../preprint}
%\bibliography{../Bib/QuantGra}

\begin{thebibliography}{10}

\bibitem{NonSingSol}
J.~M.~M.\ Senovilla,
\newblock New Class of Inhomogeneous Cosmological Perfect-Fluid Solutions
  without Big-Bang Singularity,
\newblock {\em Phys.\ Rev.\ Lett.} 64 (1990) 2219--2221

\bibitem{Sing}
M.\ Bojowald,
\newblock Absence of a Singularity in Loop Quantum Cosmology,
\newblock {\em Phys.\ Rev.\ Lett.} 86 (2001) 5227--5230, [gr-qc/0102069]

\bibitem{HomCosmo}
M.\ Bojowald,
\newblock Homogeneous loop quantum cosmology,
\newblock {\em Class.\ Quantum Grav.} 20 (2003) 2595--2615, [gr-qc/0303073]

\bibitem{LivRev}
M.\ Bojowald,
\newblock Loop Quantum Cosmology,
\newblock {\em Living Rev.\ Relativity} 11 (2008) 4, [gr-qc/0601085],
\newblock {\tt http://www.livingreviews.org/lrr-2008-4}

\bibitem{Springer}
M.\ Bojowald,
\newblock {\em Quantum Cosmology: A Fundamental Theory of the Universe},
\newblock Springer, New York, 2011

\bibitem{cosmoIV}
M.\ Bojowald,
\newblock Loop Quantum Cosmology IV: Discrete Time Evolution,
\newblock {\em Class.\ Quantum Grav.} 18 (2001) 1071--1088, [gr-qc/0008053]

\bibitem{GenericBounce}
G.\ Date and G.~M.\ Hossain,
\newblock Genericity of Big Bounce in isotropic loop quantum cosmology,
\newblock {\em Phys.\ Rev.\ Lett.} 94 (2005) 011302, [gr-qc/0407074]

\bibitem{QuantumBigBang}
A.\ Ashtekar, T.\ Pawlowski, and P.\ Singh,
\newblock Quantum Nature of the Big Bang,
\newblock {\em Phys.\ Rev.\ Lett.} 96 (2006) 141301, [gr-qc/0602086]

\bibitem{BounceReview}
M.\ Novello and S.~E.~P.\ Bergliaffa,
\newblock Bouncing cosmologies,
\newblock {\em Phys.\ Rep.} 463 (2008) 127--213

\bibitem{BouncePert}
M.\ Bojowald,
\newblock Large scale effective theory for cosmological bounces,
\newblock {\em Phys.\ Rev.\ D} 75 (2007) 081301(R), [gr-qc/0608100]

\bibitem{BounceSqueezed}
M.\ Bojowald,
\newblock How quantum is the big bang?,
\newblock {\em Phys.\ Rev.\ Lett.} 100 (2008) 221301, [arXiv:0805.1192]

\bibitem{EffectiveGR}
J.~F.\ Donoghue,
\newblock General relativity as an effective field theory: The leading quantum
  corrections,
\newblock {\em Phys.\ Rev.\ D} 50 (1994) 3874--3888, [gr-qc/9405057]

\bibitem{BKL}
V.~A.\ Belinskii, I.~M.\ Khalatnikov, and E.~M.\ Lifschitz,
\newblock A general solution of the Einstein equations with a time singularity,
\newblock {\em Adv.\ Phys.} 13 (1982) 639--667

\bibitem{PastAttract}
C.\ Uggla, H.\ van Elst, J.\ Wainwright, and G.~F.~R.\ Ellis,
\newblock The past attractor in inhomogeneous cosmology,
\newblock {\em Phys.\ Rev.\ D} 68 (2003) 103502, [gr-qc/0304002]

\bibitem{GeneralHarmonic}
D.\ Garfinkle,
\newblock Harmonic coordinate method for simulating generic singularities,
\newblock {\em Phys.\ Rev.\ D} 65 (2002) 044029, [gr-qc/0110013]

\bibitem{WS:AR}
A.~D.\ Rendall,
\newblock The nature of spacetime singularities, In A.\ Ashtekar, editor, {\em
  100 Years of Relativity -- Space-Time Structure: Einstein and Beyond},
\newblock World Scientific, Singapore, 2005, [gr-qc/0503112]

\bibitem{DiracHamGR}
P.~A.~M.\ Dirac,
\newblock The theory of gravitation in Hamiltonian form,
\newblock {\em Proc.\ Roy.\ Soc.\ A} 246 (1958) 333--343

\bibitem{Regained}
S.~A.\ Hojman, K.\ Kucha\v{r}, and C.\ Teitelboim,
\newblock Geometrodynamics Regained,
\newblock {\em Ann.\ Phys.\ (New York)} 96 (1976) 88--135

\bibitem{LagrangianRegained}
K.~V.\ Kucha\v{r},
\newblock Geometrodynamics regained: A Lagrangian approach,
\newblock {\em J.\ Math.\ Phys.} 15 (1974) 708--715

\bibitem{Rov}
C.\ Rovelli,
\newblock {\em Quantum Gravity},
\newblock Cambridge University Press, Cambridge, UK, 2004

\bibitem{ALRev}
A.\ Ashtekar and J.\ Lewandowski,
\newblock Background independent quantum gravity: A status report,
\newblock {\em Class.\ Quantum Grav.} 21 (2004) R53--R152, [gr-qc/0404018]

\bibitem{ThomasRev}
T.\ Thiemann,
\newblock {\em Introduction to Modern Canonical Quantum General Relativity},
\newblock Cambridge University Press, Cambridge, UK, 2007, [gr-qc/0110034]

\bibitem{AreaVol}
C.\ Rovelli and L.\ Smolin,
\newblock Discreteness of Area and Volume in Quantum Gravity,
\newblock {\em Nucl.\ Phys.\ B} 442 (1995) 593--619, [gr-qc/9411005],
\newblock Erratum: {\em Nucl.\ Phys.\ B} 456 (1995) 753

\bibitem{EffAc}
M.\ Bojowald and A.\ Skirzewski,
\newblock Effective Equations of Motion for Quantum Systems,
\newblock {\em Rev.\ Math.\ Phys.} 18 (2006) 713--745, [math-ph/0511043]

\bibitem{Karpacz}
M.\ Bojowald and A.\ Skirzewski,
\newblock Quantum Gravity and Higher Curvature Actions,
\newblock {\em Int.\ J.\ Geom.\ Meth.\ Mod.\ Phys.} 4 (2007) 25--52,
  [hep-th/0606232].

\bibitem{LTBII}
M.\ Bojowald, J.~D.\ Reyes, and R.\ Tibrewala,
\newblock Non-marginal LTB-like models with inverse triad corrections from loop
  quantum gravity,
\newblock {\em Phys.\ Rev.\ D} 80 (2009) 084002, [arXiv:0906.4767]

\bibitem{JR}
J.~D.\ Reyes,
\newblock {\em Spherically Symmetric Loop Quantum Gravity: Connections to
  2-Dimensional Models and Applications to Gravitational Collapse},
\newblock PhD thesis, The Pennsylvania State University, 2009

\bibitem{ThreeDeform}
A.\ Perez and D.\ Pranzetti,
\newblock On the regularization of the constraints algebra of Quantum Gravity
  in $2+1$ dimensions with non-vanishing cosmological constant,
\newblock {\em Class.\ Quantum Grav.} 27 (2010) 145009, [arXiv:1001.3292]

\bibitem{ScalarHol}
T.\ Cailleteau, J.\ Mielczarek, A.\ Barrau, and J.\ Grain,
\newblock Anomaly-free scalar perturbations with holonomy corrections in loop
  quantum cosmology, [arXiv:1111.3535]

\bibitem{QSDI}
T.\ Thiemann,
\newblock Quantum Spin Dynamics {(QSD)},
\newblock {\em Class.\ Quantum Grav.} 15 (1998) 839--873, [gr-qc/9606089]

\bibitem{InvScale}
M.\ Bojowald,
\newblock Inverse Scale Factor in Isotropic Quantum Geometry,
\newblock {\em Phys.\ Rev.\ D} 64 (2001) 084018, [gr-qc/0105067]

\bibitem{ModCollapse}
A.\ Kreienbuehl, V.\ Husain, and S.~S.\ Seahra,
\newblock Modified general relativity as a model for quantum gravitational
  collapse,
\newblock {\em Class.\ Quantum Grav.} 29 (2012) 095008, [arXiv:1011.2381]

\bibitem{TwoPlusOneDef}
A.\ Henderson, A.\ Laddha, and C.\ Tomlin,
\newblock Constraint algebra in LQG reloaded : Toy model of a ${\rm U}(1)^{3}$
  Gauge Theory I, [arXiv:1204.0211]

\bibitem{ConstraintAlgebra}
M.\ Bojowald, G.\ Hossain, M.\ Kagan, and S.\ Shankaranarayanan,
\newblock Anomaly freedom in perturbative loop quantum gravity,
\newblock {\em Phys.\ Rev.\ D} 78 (2008) 063547, [arXiv:0806.3929]

\bibitem{BohmEuclidean}
N.\ Pinto-Neto and E.~S.\ Santini,
\newblock Must Quantum Spacetimes Be Euclidean?,
\newblock {\em Phys.\ Rev.\ D} 59 (1999) 123517, [arXiv:gr-qc/9811067]

\bibitem{Action}
M.\ Bojowald and G.~M.\ Paily,
\newblock Deformed General Relativity and Effective Actions from Loop Quantum
  Gravity, [arXiv:1112.1899]

\bibitem{DegFull}
M.\ Bojowald,
\newblock Degenerate Configurations, Singularities and the Non-Abelian Nature
  of Loop Quantum Gravity,
\newblock {\em Class.\ Quantum Grav.} 23 (2006) 987--1008, [gr-qc/0508118]

\end{thebibliography}

\end{document}